\documentclass[aps,pra]{revtex4}
\begin{document}
\voffset=.5in

\title
{Spin and Statistics in Galilean Covariant Field Theory}

\author {C. R. Hagen\cite{Hagen}}

\affiliation {Department of Physics and Astronomy\\ University of
Rochester\\ Rochester, N.Y. 14627}

\begin{abstract}

The existence of a possible connection between spin and statistics
is explored within the framework of Galilean covariant field
theory.  To this end fields of arbitrary spin are constructed and
admissible interaction terms introduced. By explicitly solving
such a model in the two particle sector it is shown that no spin
and statistics connection can be established.

\end{abstract}

\maketitle
\bigskip

\noindent {\bf I. Introduction}

The possibility of a connection between spin and statistics was
first explored by Pauli \cite{Pauli}  who considered the problem
within the framework of relativistic quantum field theory.  He
showed that integer spin fields must satisfy Bose-Einstein
statistics and that half-integral fields must satisfy Fermi-Dirac
statistics provided that no other (i.e., parastatistics)
possibilities are considered.  Crucial to his proof are the
requirements that the energy be positive and that observables
commute for space-like separations.  Since both of these concepts
have no counterparts in the nonrelativistic theory \cite{concept},
it has reasonably been concluded that there is no spin and
statistics connection for the nonrelativistic case.  This has been
confirmed by the analysis of Messiah and Greenberg \cite{MG} who
showed that in nonrelativistic quantum mechanics both symmetric
and antisymmetric wave functions are allowable for any spin.

However, this conclusion has been challenged by Peshkin \cite{P1}
who claims to have established the need for symmetric wave
functions in the spin zero case.  This in turn has been rebutted
by Allen and Mondragon \cite{AM} and counter-rebutted by Peshkin
\cite{P2}. More recently Shaji and Sudarshan \cite{SS} have
claimed to establish the usual spin and statistics connection in
the nonrelativistic theory provided that the underlying Lagrangian
has a certain symmetry property.

The present paper is motivated by this continuing debate which
seemingly indicates a certain reluctance to be limited by formal
analyses such as those of ref. \cite{MG}.  It is based on two
simple observations. i) The most restrictive group of space-time
transformations which can be viewed as a limit of special
relativity is that of the Galilei group.  This leads one to the
conclusion that if a connection between spin and statistics is to
exist anywhere outside the realm of relativistic quantum field
theory, it is to be found in the corresponding Galilean limit. ii)
If nontrivial counterexamples to the usual spin and statistics
connection can be found, it immediately becomes clear that no
claim to a general connection can be tenable.

In the following section the spin one-half Galilean wave equation
is reviewed together with its extension to the arbitrary spin case
by means of the multispinor formalism.  This is used in section
 {\bf  III} to construct possible interaction terms.  The resulting
system is solved in the two particle sector to obtain a bound
state solution as well as the two particle scattering phase shift.
It is found that the solution is consistent independent of the
type of statistics satisfied by the particle (or field).

\bigskip

\noindent {\bf II. Arbitrary Spin Galilean Fields}

In principle the wave equation for a particle of arbitrary spin
can be found by using the spin one-half equation as the basic
ingredient in a multispinor approach.  In particular if one takes
a rank $2s$ multispinor which is symmetric in all its indices one
can expect to obtain an equation for a particle of spin $s$.  This
approach is totally successful and has in fact been used by Hurley
and the author \cite{HH} to derive the $g$-factor of an arbitrary
spin particle.

To carry out the indicated program one begins with the spin
one-half Galilean wave equation.  This has been derived by
L\'evy-Leblond \cite{L1} some years ago and a brief review of his
result is essential as a preliminary to its application to the
multispinor approach.  In units in which $\hbar=1$ the relevant
wave equation for a particle of mass $m$ is derived from the
Lagrangian
\[
    {\cal L}= \psi^\dagger G\psi
    \]
where
\begin{equation}
G=(1/2)(1+\rho_3)\left[ i{\partial \over \partial t}-U_0 \right] +
i\rho_1\mbox{\boldmath $\sigma$} \cdot \mbox{\boldmath $\nabla$} +
m(1-\rho_3)
\end{equation}
\noindent where two commuting sets of Pauli matrices $\rho_i$ and
$\sigma_i$ have been used to span the $4\times 4$ dimensional
spinor space.  The quantity $U_0$ which appears in conjunction
with the time derivative term is the so-called internal energy
term.

Under the Galilean transformation
\begin{eqnarray}
{\bf x^\prime} & = &  R{\bf x} +{\bf v} +{\bf a} \nonumber \\
t^\prime  & = &  t+b
\end{eqnarray}
the wave function $\psi$ transforms as $$\psi^\prime ({\bf
x^\prime}, t^\prime)= \exp [if ({\bf x},t)] \Delta^{{1\over
2}}({\bf v},R) \psi ({\bf x},t)$$ where $$f({\bf x},t) = {1\over
2} mv^2 t+m{\bf v}\cdot R{\bf x}$$ and $$\Delta^{{1\over 2}}({\bf
v},R) = \left(
\begin{array}{lr}
1  & 0  \\ -{1\over 2} \mbox{\boldmath $\sigma$} \cdot {\bf v}  &
1
\end{array}
\right) D^{{1\over 2}} (R) .$$ Note that $D^{{1\over 2}}(R)$ is
the usual two-dimensional (i.e., spin one-half) representation of
rotations which acts in the space of the $\sigma$ matrices.  It is
clear from the form of $\Delta^{{1\over 2}}({\bf v},R)$ that the
upper components of $\psi$ do not mix with the lower components
under a Galilean transformation, an important consequence of which
is the fact that the matrix $$\Gamma={1\over 2}(1+\rho_3)$$ is
invariant under all Galilean transformations.

Upon writing $\psi$ in terms of the two-component spinors $\phi$
and $\chi$ $$\psi= \left(
\begin{array}{c}
 \phi \\ \chi
\end{array}
\right) $$ one can write the Lagrangian equation $G\psi=0$ as
$$E\phi+\mbox{\boldmath $\sigma$} \cdot {\bf p}\chi=U_0\phi$$
$$\mbox{\boldmath $\sigma$} \cdot {\bf p}\phi+2m\chi=0$$ where
$E=i{\partial \over \partial t}$ and ${\bf p} = -i\mbox{\boldmath
$\nabla$}$. These equations are combined to yield the free
Schr\"{o}dinger equation $$\left( E- { {\bf p}^2\over 2m} \phi
\right) = U_0\phi .$$

One now proceeds to construct the spin $s$ wave function as a
completely symmetrized $2s$-rank spinor $\psi_{a_1...a_{2s}}$
where each $a_i$ ranges from 1 to 4.  In the absence of additional
restrictions such an object has ${1\over 6}(2s+3)(2s+2)(2s+1)$
components.  Using the fact that $\Gamma$ is an invariant matrix
one readily sees that an appropriate multispinor extension of the
spin one-half Lagrangian is given by
\[
{\cal L} = {1\over 2s}\psi^{\dagger}_{a_1...a_{2s}} \left[
\sum_{i=1}^{2s}
\Gamma_{a_1a^\prime_1}...\Gamma_{a_{i-1}a^\prime_{i-1}} G_{a_i
a^\prime_i}
\Gamma_{a_{i+1}a^\prime_{i+1}}...\Gamma_{a_{2s}a^\prime_{2s}}
\right] \psi_{a^\prime_1 ...a^\prime_{2s}}.
\]
The resulting wave equation is
\begin{equation}
\sum_{i=1}^{2s}\Gamma_{a_1a^\prime_1}...\Gamma_{a_{i-1}a^\prime_{i-1}}
G_{a_ia^\prime_i}\Gamma_{a_{i+1}a^\prime_{i+1}}...
\Gamma_{a_{2s}a^\prime_{2s}} \psi_{a^\prime_1...a^\prime_{2s}}=0.
\end{equation}
In the case of special relativity it is not possible to remove the
summation in Eq.(3) to obtain standard Bargmann-Wigner equations.
However, in Galilean relativity it can be shown \cite{CRH} that
the summation can be eliminated to obtain
\begin{equation}
\Gamma_{a_1a^\prime_1}...\Gamma_{a_{i-1}a^\prime_{i-1}}G_{a_ia^\prime_i}
\Gamma_{a_{i+1}a^\prime_{i+1}}
...\Gamma_{a_{2s}a^\prime_{2s}}\psi_{a^\prime_1...a^\prime_{2s}}=0.
\end{equation}
Thus a Bargmann-Wigner set of equations has been obtained.

Because the matrix $\Gamma$ projects out only upper components,
the presence of $(2s-1)$ $\Gamma$ matrices in (4) means that those
components of $\psi$ in which more than one index is a 3 or 4 drop
out of the equations of motion.  One thus defines the $2s+1$
components $$\psi_{a_1...a_{2s}}= \phi_{a_1...a_{2s}}$$ for
$a_i=1,2$ and the $4s$ components $$\psi_{a_1...a_{2s-1}r} =
\chi_{a_1...a_{2s-1}}^{r-2}$$ for $a_i = 1,2; r = 3,4$ to obtain
\begin{equation}
(E-U_0)\phi_{a_1...a_{2s}}+{1\over 2s} \left[ \sum_{i=1}^{2s}
\mbox{\boldmath $\sigma$}_{a_ir} \cdot{\bf p} \right]
\chi_{a_1...a_{i-1}a_{i+1}...a_{2s}}^r  =0
\end{equation}
and
\begin{equation}
\mbox{\boldmath $\sigma$}_{ra_{2s}}\cdot{\bf p}\phi_{a_1...a_{2s}}
+2m\chi_{a_1...a_{2s-1}}^r = 0.
\end{equation}
Upon solving (6) for $\chi_{a_1...a_{2s-1}}^r$ and inserting the
result into (5) one obtains the free Schr\"{o}dinger equation for
the independent components $\phi_{a_1...a_{2s}}$ $$(E-U_0-{{\bf
p}^2\over 2m}) \\  \phi_{a_1...a_{2s}}=0.$$

\bigskip

\noindent {\bf III. Interacting Arbitrary Spin Fields}

As is well known Galilean field theory is characterized by the
existence of a central extension of the algebra of the Galilei
group. It is designated as the mass operator and (in the context
of the preceding section) is given by \cite{QFT} $$M=\int d^3x
\phi^\dagger_{a_1...a_{2s}} \phi_{a_1...a_{2s}}.$$ The operator
$M$ has the property that it commutes with all the operators of
the Galilei group and implies the so-called Bargmann
superselection rule.  The latter has the consequence that a given
Galilean field theory factors into sectors labelled by the
eigenvalue of $M$.  Thus the vacuum state is the state which is
the null eigenvalue of $M$ and satisfies
$$\phi_{a_1...a_{2s}}|0\rangle =0.$$ This has the consequence that
the usual two-point function is no longer a time ordered function
but has the degenerate form $$G_{a_1...a_{2s};
a^\prime_1...a^\prime_{2s}} ( {\bf x}  -  {\bf
x^\prime},t-t^\prime ) = -i  \theta (t-t^\prime) \langle
0|\phi_{a_1...a_{2s}}({\bf x},t)
\phi^\dagger_{a^\prime_1...a^\prime_{2s}} ({\bf
x^\prime},t^\prime) | 0 \rangle.$$ This clearly implies that the
two-point function is insensitive to the statistics of the field
in question.  Alternatively stated, the one-particle state (like
the vacuum) cannot be used to test the consistency of a particular
type of statistics. Accordingly, attention is now focused on the
two particle sector of the spin $s$ theory formulated in section
{\bf II}.

The goal here is to determine whether the choice of statistics for
the $\phi$ particle is constrained in any way by the value of the
spin.  While this issue could be pursued in the free field limit,
it is clearly of greater interest to examine it for the case of a
nontrivial interaction.  One possible Galilean invariant
interaction is given by the Lagrangian \cite{L2}

\begin{equation}
{\mathcal{L}}^\prime={\lambda\over 2}{\bf J^\dagger}\cdot{\bf J}
\end{equation}
where $${\bf J}=\int d \mbox{\boldmath $\xi$} \phi_{a_1...a_{2s}}
\left( {\bf x}+{1\over 2}\mbox{\boldmath $\xi$},t \right)
\mbox{\boldmath $\nabla$} f^* (|\mbox{\boldmath $\xi$}|)
\Sigma^2_{a_1...a_{2s}; a^\prime_1...a^\prime_{2s}}
\phi_{a^\prime_1...a^\prime_{2s}} \left( {\bf x}-{1\over 2}
\mbox{\boldmath $\xi$},t \right)$$ where
$$\Sigma^2_{a_1...a_{2s};a^\prime_1...a^\prime_{2s}}
=\sigma^2_{a_1a^\prime_1}...\sigma^2_{a_{2s}a^\prime_{2s}}$$ and
$\sigma^2$ is the usual Pauli matrix.  Such an interaction is
Galilean invariant for all $f(\mbox{\boldmath $\xi$})$ which
depend only on the scalar $|\mbox{\boldmath $\xi$}|$. It is
important to note that one seeks here to impose the ``wrong"
statistics relation $$\phi_{a_1...a_{2s}} ({\bf x},t)
\phi_{a^\prime_1...a^\prime_{2s}} ({\bf x}^\prime,t) + (-1)^{2s}
\phi_{a^\prime_1...a^\prime_{2s}} ({\bf x}^\prime,t)
\phi_{a_1...a_{2s}} ({\bf x},t)=0.$$ Had one chosen to apply the
usual statistics relation a scalar-scalar coupling (i.e., no
derivative of $f(\mbox{\boldmath $\xi$}))$ rather than the
vector-vector coupling (7) would be the simplest interaction
choice. This is readily seen to be a consequence of the transpose
relation $$(\Sigma^2)^T=(-1)^{2s}\Sigma^2 .$$

Since the internal energy is irrelevant for purposes of this
discussion, it can be discarded, thereby leading to the
Hamiltonian
\begin{equation}
H={-1\over 2m}\int d{\bf x} \phi^{\dagger}_{a_1...a_{2s}}
\nabla^2\phi_{a_1...a_{2s}} -{\lambda\over 2}\int d{\bf x} {\bf
J^\dagger}\cdot{\bf J}.
\end{equation}
Thus the equation to be solved is
\begin{equation}
H|N\rangle=E_N|N\rangle
\end{equation}
where $E_N$ is the energy of the $N$-particle state. Note that the
$N$-particle state also satisfies the eigenvalue equation $$M|N
\rangle = Nm|N \rangle .$$ For the cases $N=0$ and $N=1$ Eq. (8)
is trivially solved since the interaction term is inoperative.
Thus subsequent attention is devoted entirely to the case $N=2$.

One looks for a solution of the form
\begin{equation}
|2\rangle=\int\int d{\bf x}d{\bf x^\prime}
\phi^{\dagger}_{a_1...a_{2s}} ({\bf x})
\Sigma^2_{a_1...a_{2s};a^\prime_1...a^\prime_{2s}}\phi^{\dagger}_{a^\prime_1...
a^\prime_{2s}}({\bf x^\prime})\exp[i{\bf P}\cdot({\bf x}+{\bf
x^\prime})] \psi({\bf x-x^\prime})|0\rangle
\end{equation}
corresponding to a center-of-mass momentum ${\bf P}$. This leads
to an equation of the form
\begin{equation}
\left[ E-{{\bf P}^2\over 4m}+{1\over m}\nabla^2 \right]\psi(x)
=
-\lambda2^{2s}\nabla_jf({\bf x}) \int d{\bf
x^\prime}\nabla^\prime_j f^*{\bf x} \psi({\bf x^\prime}).
\end{equation}
This is a well known example of an integral equation with a
separable kernel which by standard techniques leads to the
eigenvalue condition $$1=-\lambda{2^{2s}\over 3}\int {d {\bf
p}\over (2\pi)^3} \;\; { {\bf p}^2|f(p)|^2 \over \Omega - {{\bf
p}^2/m} }$$ where $$\Omega\equiv E-{{\bf P}^2\over 4m}$$ is the
renormalized internal energy of an assumed bound state solution.
Such a $\Omega<0$ solution will exist when the condition
$$m\lambda{2^{2s}\over 3}\int {d{\bf p}\over (2\pi)^3}|f(p)|^2>1$$
is satisfied.  Although the local limit $f(p)\to  1$ is not
allowable in a finite theory, this has no bearing on the overall
consistency of the theory since any rotationally invariant $f(p)$
is in fact consistent with Galilean invariance with ``wrong"
statistics.

Scattering solutions can also be obtained for Eq.(11).  In this
case one infers for a particle of incoming momentum ${\bf k}$ and
internal energy $U= {{\bf k}^2\over m}$ a solution of the form
$$\psi({\bf p})=(2\pi)^3 {1\over 2} [\delta ( {\bf p}-{\bf
k})-\delta ({\bf p}+{\bf k})] -\lambda 2^{2s} { p_jf(p) e^{i{\bf
p}\cdot{\bf x}} \over U -{\bf p}^2/m } \int {d {\bf q}\over
(2\pi)^3} q_jf^*(q)\psi(q)$$ which leads to the explicit result
for $\psi({\bf x})$
\begin{eqnarray*} \psi ({\bf x})  & = &
{1\over 2}(e^ {i{\bf k} \cdot{\bf x}} - e^{-i{\bf k}\cdot{\bf x}}
) -\lambda 2^{2s}
 \int {d{\bf p}\over (2\pi)^3}
e^{i{\bf p}\cdot {\bf x}} \; { {\bf p}_0\cdot{\bf p} f(p)f^*(p_0)
\over U - {{\bf p}^2/m}} \\ & & \left[ 1+\lambda{2^{2s}\over 3}
\int {d{\bf q}\over (2\pi)^3} \; { {\bf q}^2 |f(q)|^2 \over
U-{{\bf q}^2/m}} \right]^{-1}.
\end{eqnarray*}

Upon performing the integration over ${\bf p}$ this is found to
yield $$\psi({\bf x}) ={1\over 2} \left( e^{i{\bf k}\cdot{\bf x}}
-  e^{-i{\bf k}\cdot{\bf x}} \right) +m\lambda 2^{2s} \left( {\bf
k}\cdot{1\over i}{\bf \nabla}\right) {e^{ikr}\over r} \left[
1+\lambda{2^{2s}\over 3}\int {d{\bf q}\over (2\pi)^3} \; {{\bf
q}^2|f(q)|^2 \over U-{{\bf q}^2/m}}\right]^{-1},$$ thereby
displaying the fact that there is scattering only of the $P$ wave.

This allows identification of the $P$ wave phase shift $\delta_1$
as $$e^{i\delta_1}\sin\delta_1=m\lambda{2^{2s}\over 3}k^3 \left[
1+\lambda{2^{2s}\over 3} \int {d{\bf q}\over (2\pi)^3} \; {{\bf
q}^2|f(q)|^2\over U-{{\bf q}^2/m}} \right]^{-1}.$$ This in turn
allows one to establish exact equivalence to the effective range
formula appropriate to P wave scattering
$$k^3\cot\delta_1=-{1\over a}+{1\over 2}r_ok^2$$ where the
scattering length (or scattering volume) is given by $$-{1\over
a}= {4\pi \zeta \Omega\over m} -{1\over 2}(-m \Omega)^{3\over 2}$$
where $$\zeta \equiv \int {d{\bf q}\over (2\pi)^3} \; { {\bf
q}^2|f(q)|^2 \over (\Omega - q^2/m)^2} \; .$$ Similarly one finds
for $r_0$ $$r_0 =  {-8\pi\zeta\over m^2} -3(-m \Omega)^{1\over
2}.$$ In sum the Galilean spin $s$ theory is totally consistent
despite its having been quantized with ``wrong" statistics.  It
has two divergences in the local limit $f(p)=1$, the linearly
divergent $\zeta$ and the cubically divergent internal energy
$\Omega$. Although it is not feasible to extend the solution to
the general sectors of the model, it seems clear that no
complications are likely to arise in such cases.  Were such to
occur, they would lead to the conclusion that the statistics of
particle pairs are unavoidably linked to the presence of other
particles. Such possibilities are not to be found in the arguments
normally raised against theories with ``wrong" statistics,
however.

\bigskip

\noindent {\bf IV. Conclusion}

The question as to whether ``wrong" statistics field theories can
be excluded on the basis of very general arguments has been
examined here within the framework of a Galilean multispinor
formalism.  It has in fact been demonstrated that  there is no
particular difficulty in constructing such theories.  While the
imposition of conventional statistics is somewhat more natural in
the sense that they allow scalar-scalar interactions to be
accommodated, it has been demonstrated here that vector-vector
interactions pose no special difficulty.

As a final remark it should be noted that the usual symmetric
multispinor method applies only to nonzero spin--namely, one
cannot construct zero spin from a symmetric multispinor.  In that
special case it is, however, sufficient to observe that an
antisymmetric spinor suffices to derive the spinless counterpart
of Eq.(8).  The details of such an approach are to  be found in
[8].

\bigskip

This work is supported in part by the U.S. Department of Energy
Grant No. DE-FG02-91ER40685.

\newpage



\begin{thebibliography}{20}
\bibitem{concept} It may be useful to remark here that the nonrelativistic
theory requires only that the energy spectrum be bounded from
below, not that it necessarily be positive.
\end{thebibliography}
\end{document}